\documentstyle[aps,preprint,12pt]{revtex}

\begin{document}

\title{Self Consistent and Renormalized particle-particle RPA in a
  Schematic Model}

\author{E.J.V. de Passos$^\dagger$, A.F.R. de Toledo Piza$^\dagger$ and F.
Krmpoti\'c$^\ddag$} 

\address{$^\dagger$Instituto de F\'{\i}sica, Universidade de S\~{a}o
Paulo \\ Caixa Postal 66318, 05389-970 S\~{a}o Paulo, SP, Brazil}

\address{$^\ddag$Departamento de Fisica, Facultad de Ciencias
Exactas, Universidad Nacional de La Plata \\ C.C. 67, 1900 La Plata,
Argentina}

\maketitle

\begin{center}
{\today}
\end{center}

\begin{abstract}
The dynamical effects of ground state correlations for excitation
energies and transition strengths near the superfluid phase transition
are studied in the soluble two level pairing model, in the context of
the particle-particle self consistent Random Phase Approximation
(SCRPA). Exact results are well reproduced across the transition
region, beyond the collapse of the standard particle-particle Random
Phase Approximation. The effects of two-body correlation in the SCRPA
are displayed explicitly.

\end{abstract}

\vspace{3cm}


\newpage

An extension of the self-consistent Random Phase Approximation (SCRPA)
to the particle-particle case (pp-SCRPA) has been recently presented
by Dukelsky, R\"opke and Schuck \cite{one}. The main concern there is
with differences between ground state correlations due to pairing
vibrations and Bruekner-Hartree-Fock ground state correlations. In
this report we use the soluble schematic model used in \cite{one} to
study different aspects of the pp-SCRPA. Namely, we examine the
behavior of the system as we approach the transition point from the
normal to the superfluid phase and investigate the importance of the
two-body ground state correlations for excitation energies and
transition strengths in this domain.

The model is the standard two-level pairing model whose Hamiltonian is
given by

\begin{equation}
\label{um}
H = \frac{\epsilon}{2}\sum_\sigma\sigma N_\sigma-g\Omega \sum_{\sigma,
\sigma'} A^+_\sigma A_{\sigma'}
\end{equation}

\noindent where $\sigma= \pm 1$ labels the upper and lower levels
which have equal pair degeneracy $\Omega = j + 1/2$ and $N_\sigma$,
$A^+_\sigma$ are respectively the number and pair operators in the
level $\sigma$, i.e.

\begin{equation}
\label{dois}
N_\sigma = \sum_m a^+_{\sigma m} a_{\sigma m}\;, \quad \quad
A^+_{\sigma m} = \frac{1}{\sqrt{\Omega}} \frac{1}{2} \sum_m
(-)^{j-m} a^+_{\sigma m} a^+_{\sigma-m}.
\end{equation}

\noindent For each level $\sigma$ the operators

\begin{equation}
\label{tres}
S_+^{(\sigma)} = \sqrt{\Omega} A^+_\sigma\;, \quad S_-^{(\sigma)} =
S_+{^{(\sigma)}}^+\;, \quad S_0^{(\sigma)} = \frac{1}{2} \left(
N_\sigma - \Omega \right)
\end{equation}

\noindent satisfy an SU(2) algebra.

The addition, $P^+$, and removal, $R^+$, phonon creation operators of
the pp-SCRPA for the model are \cite{one}
  
\begin{equation}
\label{quatrocinco}
P^+  =  \lambda A^+_1 - \mu A^+_{-1}\;, \quad 
R^+  =  \rho A_{-1} - \tau A_1  
\end{equation}

\noindent and the pp-SCRPA ground state $\mid 0\rangle$ is defined as
the vacuum of these phonons, i.e.

\begin{equation}
\label{oito}
P \mid 0 \rangle = R \mid 0 \rangle = 0.
\end{equation}

\noindent Addition and removal excitations are normalized as

\begin{equation}
\label{seis}
\langle 0| [P,P^+] | 0 \rangle = 1\;, \quad
 \langle 0| [R,R^+] | 0 \rangle = 1\;, \quad
  \langle 0| [R^+,P^+] | 0 \rangle = 0 
\end{equation}

\noindent which gives, using Eqs. (\ref{quatrocinco}),

\begin{eqnarray}
&&\lambda^2 \langle 0 | \left( 1 - \frac{N_1}{\Omega} \right)
  | 0 \rangle + \mu^2 \, \langle 0 | \left( 1 - \frac{N_{-1}}{\Omega}
   \right) |0 \rangle = 1  \nonumber \\
\label{sete}
&&\tau^2 \langle 0 | \left( 1 - \frac{N_1}{\Omega} \right)
  | 0 \rangle + \rho^2 \, \langle 0 | \left( 1 - \frac{N_{-1}}{\Omega}
   \right) |0 \rangle = - 1  \\
&&\tau\lambda \langle 0 | \left( 1 - \frac{N_1}{\Omega} \right)
  | 0 \rangle + \mu\rho \, \langle 0 | \left( 1 - \frac{N_{-1}}{\Omega}
   \right) |0 \rangle = 0. \nonumber
\end{eqnarray}

The pp-SCRPA equations for the present model can be written in the
form \cite{one,two}

\begin{equation}
\label{nove}
\left( \begin{array}{ll} A & B \\ B^* & C \end{array}\right)
\left( \begin{array}{l}  \lambda \\ \mu \end{array} \right) =  \omega
\left( \begin{array}{ll}  U & 0 \\ 0 & V \end{array}\right)
\left( \begin{array}{l}  \lambda \\ \mu \end{array}\right)
\end{equation}

\noindent where

\begin{eqnarray}
\label{dez}
&&A = \langle 0 \mid \left[ A_1, H, A^+_1 \right] \mid 0 \rangle\;, 
\quad 
B = - \langle 0 \mid \left[ A_1, H, A^+_{-1}\right]\mid 0\rangle\;, 
\nonumber \\
&&C =\langle 0\mid \left[ A_{-1},H, A^+_{-1}\right]\mid 0\rangle\;,
\quad
U= \langle 0 \mid \left[ A_1, A^+_1 \right] \mid 0 \rangle\;, \\ 
&&V = \langle 0 \mid\left[ A_{-1}, A^+_{-1}\right] \mid 0\rangle
\nonumber 
\end{eqnarray}

\noindent with the usual definition of the symmetrized double
commutator 

\[
[A,B,C] = \frac{1}{2} \left( [A,[B,C]] + [[A,B], C] \right).
\]

Equation(\ref{nove}) has two solutions. We associate the solution of
positive norm (normalized to +1) with the addition phonon, $\lambda_+
= \lambda, \; \mu_+ = \mu$ and the corresponding eigenfrequency with
$\omega_+ = E_0(A+2) - E_0(A)$. The negative norm solution (normalized
to -1) we associate with the removal phonon $\lambda_- = \tau, \;
\mu_- = \rho$ and the corresponding eigenfrequency with $\omega_- =
E_0(A) - E_0 (A-2)$ \cite{three}. This is the physical meaning of the
first two Eqs. (\ref{sete}). The last one expresses the orthogonality
of these two solutions.

>From Eqs.(\ref{sete}) we get $\tau$ and $\rho$ in terms of $\lambda$
and $\mu$ and, choosing phases to agree with ref. \cite{one}, one has

\begin{eqnarray*}
P^+ & = & \lambda A^+_1 - \mu  A^+_{-1} \\[0.3cm] R^+ & = &
\left( \frac{\langle 0 \mid (N_{-1} - \Omega) \mid 0\rangle} { \langle
0\mid \Omega - N_1 \mid0\rangle} \right)^{1/2} \mu A_1  - 
\left( \frac{\langle 0 \mid \Omega - N_1 \mid 0 \rangle}{\langle 0\mid
N_{-1} - \Omega \mid 0 \rangle} \right)^{1/2} \lambda A_{-1}.
\end{eqnarray*}

The model Hamiltonian (\ref{um}) commutes with $S{^{(1)}}^2$,
$S{^{(-1)}}^2$ and $S_0^{(1)} + S_0^{(-1)}$, where $S{^{(\sigma)}}^2 =
{\displaystyle \frac{1}{2}} \left( S_+^{(\sigma)} S_-^{(\sigma)} +
S_-^{(\sigma)} S_+^{(\sigma)} \right) + S_0{^{(\sigma)}}^2$. In
particular, the ground states of even nuclei are defined in the SU(2)
$\otimes$ SU(2) irreducible representation $\left( {\displaystyle
\frac{\Omega}{2},\frac{\Omega}{2}} \right)$. A basis of states
in this subspace is given by

\begin{equation}
\label{onze}
\mid m_1; m_2 \rangle = \left| \frac{\Omega}{2} \, m_1 \right\rangle_1 
    \otimes \left| \frac{\Omega}{2} \, m_2 \right\rangle_{-1} \ \ .
\end{equation}

\noindent The pp-SCRPA ground state of the $N =2\Omega$ system can
therefore be expanded as

\[
\mid 0 \rangle =  \sum^{\Omega}_{m=0} C_{m;-m} \left| -  
\frac{\Omega}{2} + m,\frac{\Omega}{2} - m \right\rangle
\]

\noindent where $m$ is the number of particle pairs coupled to $J=0$
in the upper-level. 

The condition that $\mid 0\rangle$ is the vacuum of the addition
phonon, Eq. (\ref{oito}) gives the recursion relation

\begin{equation}
\label{doze}
C_{m; -m} = \frac{\mu}{\lambda} \, C_{m-1; -(m-1)}
\end{equation}

\noindent which allows for writing the normalized ground state
explicitly as \cite{one}

\begin{equation}
\label{treze}
\mid 0 \rangle = {\frac{1}{\sqrt{{\displaystyle \sum^\Omega_{m=0}
\left( \frac{\mu}{\lambda} \right)^{2m}}}}}
\sum^\Omega_{m=0} \left( \frac{\mu}{\lambda} \right)^m \left| 
 - \frac{\Omega}{2} + m, \frac{\Omega}{2} - m \right\rangle.
\end{equation}

Apparently we have a problem here, since the pp-SCRPA ground state has
been determined by just one of Eqs. (\ref{oito}) and we still have the
additional condition that $\mid 0 \rangle$ is also supposed to be the
vacuum of the removal phonon. However, the second Eq. (\ref{oito})
gives for the coefficients of $\mid 0 \rangle$ the condition

\begin{equation}
\label{quatorze}
C_{m;-m} = \frac{\langle 0 \mid (N_{-1} - \Omega)
      \mid 0 \rangle}{\langle 0 \mid \Omega - N_1 \mid 0 \rangle}
        \frac{\mu}{\lambda} C_{m-1; - (m-1)}.
\end{equation}

\noindent This is consistent with Eq. (\ref{doze}) only if

\begin{equation}
\label{quinze}
\langle 0 \mid N_{-1} - \Omega \mid 0 \rangle = 
\langle 0 \mid \Omega - N_1 \mid 0 \rangle
\end{equation}

\noindent and that this is true follows from

\[
\langle 0 \mid N_1 + N_{-1} - 2\Omega \mid 0 \rangle = 0
\]

\noindent which can be interpreted as saying that the number of
particles in the level $+1$ is equal to the number of holes in
the level $-1$. Note that the above discussion uses the condition
$N=2\Omega$ (or $S_0^{(1)}+S_0^{(-1)}=0$) in an essential way.

Defining the normalized amplitudes

\[
\overline{\lambda} = \lambda \sqrt{\langle 0 \mid \left( 1 -
\frac{N_1}{\Omega}\right) \mid 0 \rangle}, \quad 
\overline{\mu} = \mu\sqrt{\langle 0 \mid \left( 1 -
\frac{N_1}{\Omega}\right) \mid 0 \rangle},
\]

\noindent with $\overline{\lambda}^2 - \overline{\mu}^2 = 1$, we can
write the pp-SCRPA equations as

\begin{equation}
\label{desesseis}
\left( \begin{array}{ll}
\overline{A} & \overline B \\ {\overline{B}}^* & \overline{C}
\end{array}\right)
\left( \begin{array}{l}\overline{\lambda} \\  \overline{\mu} 
\end{array}\right)  = \omega
\left( \begin{array}{cc} 1 & 0 \\ 0 & -1 \end{array}\right)
\left(\begin{array}{ll} \overline{\lambda} \\  \overline{\mu}
\end{array}\right)
\end{equation}

\noindent where $\overline{A}$, $\overline{B}$ and $\overline{C}$ are
now given by Eqs. (\ref{dez}) in terms of the normalized operators
${\overline{A}}^+_1$, ${\overline{A}}^+_{-1}$ defined as 

\[
{\overline{A}}^+_1 =\frac{A^+_1}{\sqrt{\langle 0 \mid \left( 1 -
\frac{N_1}{\Omega}\right)\mid 0 \rangle}}, \quad 
{\overline{A}}^+_{-1} = \frac{A^+_{-1}}{\sqrt{\langle 0 \mid \left( 1
- \frac{N_1}{\Omega}\right) \mid 0 \rangle}}.
\]

\noindent Evaluating the double commutators one obtains

\begin{eqnarray}
\label{desessete}
\overline{A} & = & \epsilon - g \frac{\langle 0\mid\left(\Omega - N_1
\right)^2\mid 0 \rangle}{\langle 0\mid \Omega - N_1 \mid 0 \rangle} +
g\langle 0 \mid 2 {\overline{A}}^+_1 \overline{A}_1 +
{\overline{A}}^+_{-1}\overline{A}_1 +\overline{A}^+_1\overline{A}_{-1}
\mid 0 \rangle \\
\label{desoito}
\overline{B} & = & g\frac{\langle 0\mid (\Omega - N_1)(\Omega -
N_{-1})\mid 0 \rangle}{\langle 0\mid \Omega - N_1\mid 0\rangle} \\
\label{desenove}
\overline{C} & = & \epsilon - g \frac{\langle 0\mid \left(\Omega -
N_{-1} \right)^2\mid 0 \rangle}{\langle 0\mid \Omega - N_1\mid 0
\rangle}+ g\langle 0\mid 2 {\overline{A}}^+_{-1} \overline{A}_{-1} +
{\overline{A}}^+_1\overline{A}_{-1} +
{\overline{A}}^+_{-1}\overline{A}_1\mid 0 \rangle. 
\end{eqnarray}

In order to calculate the expectation values appearing in these
expressions we invert Eqs. (\ref{quatrocinco}) to find
${\overline{A}}^+_1 = \overline{\lambda} P^+ - \overline\mu R$ and
${\overline{A}}^+_{-1} = \overline{\mu} P^+ - \overline\lambda R$.
>From this we then obtain, using the vacuum condition (\ref{oito}),
$\langle 0 \mid {\overline{A}}^+_1\overline{A}_1\mid 0 \rangle =
\overline{\mu}^2$, $\langle 0 \mid {\overline{A}}^+_{-1}
\overline{A}_{-1} \mid 0 \rangle = \overline{\lambda}^2$ and $\langle
0 \mid A^+_{-1} A_1 \mid 0 \rangle = \overline{\lambda}\overline{\mu}$
\cite{one}. The matrices $\overline{A}$, $\overline{B}$ and
$\overline{C}$ are thus functions of the pp-RPA amplitudes
$\overline{\lambda}$ and $\overline{\mu}$, of the one-body densities
$\langle 0\mid N_1\mid 0 \rangle$, $\langle 0\mid N_{-1}\mid 0
\rangle$ and of the two-body densities $\langle 0\mid N^2_1 - N_1\mid
0 \rangle$, $\langle 0\mid N^2_{-1} - N_{-1}\mid 0 \rangle$ and
$\langle 0 | N_1 \, N_{-1} | 0 \rangle$. Their final expressions are

\begin{eqnarray}
\label{vinte}
\overline{A} & = & \epsilon - g \, \frac{\langle 0 | \left(\Omega - N_1
   \right)^2 | 0 \rangle}{\langle 0| (\Omega - N_1) | 0 \rangle} 
    + 2g \left( \overline{\mu}\,^2 + \overline{\lambda}
     \, \overline{\mu}  \right) \\
\label{vinteum}
\overline{B} & = & g \, \frac{\langle 0 | (\Omega - N_1)(\Omega -
    N_{-1})|0 \rangle}{\langle 0| \Omega - N_1 |0\rangle}
    \\
\label{vintedois}
\overline{C} & = & \epsilon - g \, \frac{\langle 0 | \left(\Omega -
  N_{-1} \right)^2 | 0 \rangle}{\langle 0| (\Omega - N_1) | 0 \rangle}
   + 2g \left( \overline{\lambda}\,^2 + 
    \overline{\lambda} \, \overline{\mu} \right)  \ .
\end{eqnarray}

\noindent Once we have the explicit expression (\ref{treze}) for the
correlated ground state $\mid 0\rangle$, we can calculate the one- and
two-body densities and solve the non-linear pp-SCRPA equations,
Eqs. (\ref{desesseis}) and (\ref{vinte})-(\ref{vintedois}), for the
pp-RPA amplitudes $\overline{\lambda}$ and $\overline{\mu}$. Results
of such a calculation will be presented further on. Note that in the
limit of standard pp-RPA, $\mid 0\rangle$ is replaced by the
uncorrelated ground state and Eq. (\ref{desesseis}) reduces to the
ordinary equation

\begin{equation}
\label{vintetres}
\left(\begin{array}{cc}\epsilon - g\Omega & - g\Omega  \\
 - g\Omega & \epsilon + 2g - g\Omega \end{array}\right)
\left(\begin{array}{l}\overline{\lambda} \\ \overline{\mu}
\end{array}\right) = \omega 
\left(\begin{array}{cc} 1 & 0  \\  0 & -1\end{array}\right)
\left(\begin{array}{l}\overline{\lambda} \\ \overline{\mu} 
\end{array}\right).
\end{equation}
 
This illustrates the fact that, in practice, solving the SCRPA
equations is rather demanding from a computational point of view, and
in fact this has not been done so far in realistic cases. Several
authors have therefore suggested alternate schemes to derive simpler
equations which are more easily amenable to numerical treatment
\cite{two},\cite{four}-\cite{seven}. In essence, the simplifying
hypothesis consists in neglecting the two-body correlations in the
pp-SCRPA equations of motion (\ref{desesseis}). These approximations
are generically called renormalized RPA (RRPA). The pp-RRPA simplified
version of the ingredients involved in
Eqs. (\ref{desessete})-(\ref{desenove}) reads

\begin{eqnarray*}
&& \langle 0\mid N^2_1\mid 0 \rangle\simeq\langle 0\mid N_1\mid 0
\rangle^2 + \left( 1 - \frac{\langle 0\mid N_1\mid 0\rangle}{2\Omega}
\right)\langle 0\mid N_1\mid 0 \rangle \\
&& \langle 0\mid N^2_{-1}\mid 0 \rangle\simeq\langle 0\mid N_{-1}\mid
0\rangle^2+\left(1-\frac{\langle 0\mid N_{-1}\mid 0\rangle}{2\Omega}
\right)\langle 0\mid N_{-1}\mid 0 \rangle \\
&& \langle 0\mid N_1 N_{-1}\mid 0 \rangle\simeq\langle 0\mid N_1\mid 0
\rangle\langle 0\mid N_{-1}\mid 0 \rangle \\
&& \langle 0\mid A^+_1 A_1\mid 0\rangle\simeq \frac{1}{4\Omega^2}
\langle 0\mid N_1\mid 0\rangle^2 \\
&& \langle 0\mid A^+_{-1} A_{-1}\mid 0\rangle\simeq
\frac{1}{4\Omega^2}\langle 0\mid N_{-1}\mid 0 \rangle^2\\
&& \langle 0\mid A^+_{-1} A_1\mid 0 \rangle = \langle 0\mid A^+_1
A_{-1}\mid 0 \rangle\simeq 0
\end{eqnarray*}

\noindent and $\overline{A}$, $\overline{B}$ and $\overline C$ become
respectively

\begin{eqnarray*}
\overline{A} & \simeq & \epsilon - g\frac{\langle 0\mid N_1 \mid
0\rangle}{\Omega} - g \Omega\langle 0\mid \left(1- \frac{N_1}{\Omega}
\right)\mid 0 \rangle \\ \overline{B} & \simeq & - g\Omega \langle 0
\mid \left(1 - \frac{N_1}{\Omega} \right)\mid 0 \rangle \\
\overline{C} & \simeq & \epsilon + 2g - g \frac{\langle 0\mid N_1 \mid
0 \rangle}{\Omega} - g\Omega \langle 0\mid \left(1 -
\frac{N_1}{\Omega} \right)\mid 0 \rangle.
\end{eqnarray*}

The pp-RRPA equation is thus analogous to the pp-RPA equation
(\ref{vintetres}), except that now $g$ is renormalized by
${\displaystyle \left(1 - \frac{\langle 0\mid N_1\mid
0\rangle}{\Omega} \right)}$ and the particle and hole energies are
shifted respectively by $-g{\displaystyle \frac{\langle 0\mid N_1\mid
0\rangle}{2\Omega}}$ and ${g\displaystyle \frac{\langle 0\mid N_1\mid
0\rangle}{2\Omega}}$. In realistic cases \cite{four} we have to make
further approximations in the calculation of the one-body densities to
avoid having to solve the vacuum condition, Eq. (\ref{oito}). For the
calculations reported below we calculate $\langle 0\mid N_1\mid 0
\rangle$ using the expression Eq. (\ref{treze}) for the phonon vacuum
$\mid 0 \rangle$. The difference between our pp-SCRPA and pp-RRPA
results are thus solely due to neglecting two-body correlations in the
pp-SCRPA equations of motion.

We next present numerical results obtained for the two-level pairing
model. They correspond to a case in which twenty particles distribute
themselves in two $\Omega=10$ levels separated by $\epsilon=2$. In
Fig.~1 we plot the energies $\omega_+$ and $\omega_-$ as functions of
the coupling strength $g$. The pp-RPA collapses at $g_{\rm crit} =
{\displaystyle \frac{\epsilon}{2\Omega-1}}$, at which point the
corresponding energies become equal, $\omega_+ = \omega_- = -g_{\rm
crit}$.  Since $\omega_+$ and $\omega_-$ have the physical
interpretation of (minus) two-nucleon removal energies respectively
from the closed-shell-plus-two and from the closed-shell nuclei, this
equality signals the breakdown of shell structure. On the other hand,
neither the SCRPA nor the RRPA collapse, but instead follow closely
the exact values in the neighborhood of the phase transition
point. As $g$ increases, however, the agreement with the exact
energies deteriorates. In particular, the pp-SCRPA, and even more
markedly the pp-RRPA energies approach values that differ from the
exact strong coupling limits $\omega_+\rightarrow 0$ and
$\omega_-\rightarrow -2g$ for large values of $g$.  In Fig.~2 we plot
the occupation of the upper level in the ground state of the closed
shell nuclei. Again we see that in the neighborhood of the phase
transition point both approximations yield values close to the exact
ones, whereas the pp-RPA value diverges. However, when $g$ increases,
the pp-SCRPA value lags considerably behind the exact one, which
agrees well with the pp-RRPA result almost up to the strong coupling
limit of $\langle 0\mid N_1\mid 0 \rangle\rightarrow \Omega$.  Fig.~3
shows, also as a function of $g$, the square of the two-particle
transfer matrix element to the upper level:

\[
\mid\langle N+2\mid A^+_1\mid N \rangle\mid^2 = \overline\lambda\,^2
\left(1 - \frac{\langle 0\mid N_1\mid 0 \rangle}{\Omega} \right).
\]

\noindent We find again the same behavior, with agreement near the
phase transition point and disagreement when $g$ increases
approaching the strong coupling limit ${\displaystyle
\frac{\Omega+1}{4}}$. 

In conclusion, we find that correlations introduced by the pp-SCRPA or
by the pp-RRPA allow for following the system across of the phase
transition point associated with the collapse of the ordinary
RPA. However, it should be stressed that the pp-RRPA, which is in fact
an approximate treatment of the full pp-SCRPA performs even better
than the full theory for some observables. The pp-RRPA values for the
occupancy of the upper level agree with the exact values all the way
up to the strong coupling regime, and pp-RRPA values for the
two-particle transfer matrix elements also perform somewhat better than
the pp-SCPRA values. Only for the excitation energies are the pp-SCRPA
results better than the corresponding pp-RRPA results. A possible
perspective concerning these results could be formed by recalling the
well known fact that the standard RPA tends to overestimate
collectivity. In the present schematic model, on the other hand, the
full pp-SCRPA seems to underestimate it, whereas the ommision of
two-body correlations in the pp-RRPA gives improved occupancy and
transfer results. In general, when $g$ increases towards the strong
coupling limit, pp-SCRPA and pp-RRPA correlations are not adequate to
reproduce exact values, indicating that we should consider the system
as a pairing-deformed (superfluid) system \cite{eight}. This is
consistent with the fact that the pair-coupling model \cite{pc} is
asymptotically exact in the strong coupling limit.

\newpage

\begin{center}

{\bf Figure captions}

\end{center}

{\bf Figure 1.} Excitation energies $\omega_+$ (upper points and
curves) and $\omega_-$ as functions of the coupling strength $g$ for
the two-level pairing model with $\epsilon=2$, $\Omega=10$ and $N=20$.

{\bf Figure 2.} Number of particles in the upper level for the ground
state of the two-level pairing model. Same parameters as for Fig. 1.

{\bf Figure 3.} Absolute square of the two particle addition matrix
element to the ground state of the two-level pairing model. Same
parameters as for Fig. 1.

\end{document}